\documentclass[aps,pre,superscriptaddress,showpacs,preprint]{revtex4}
\usepackage{graphicx}
\usepackage{epsfig}
\newcommand{\be}{\begin{equation}}
\newcommand{\ee}{  \end{equation}}
\newcommand{\ba}{\begin{eqnarray}}
\newcommand{\ea}{  \end{eqnarray}}

\begin{document}
\title{Friedel Oscillations in Microwave Billiards}
\author{A.~B\"{a}cker}
\affiliation{Institut f\"{u}r Theoretische Physik, Technische
Universit\"{a}t Dresden, D-01062 Dresden, Germany}
\author{B.~Dietz}
\affiliation{Institut f\"{u}r Kernphysik, Technische
Universit\"{a}t Darmstadt, D-64289 Darmstadt, Germany}
\author{T.~Friedrich}
\affiliation{GSI Helmholtzzentrum f\"{u}r Schwerionenforschung
GmbH, D-64291 Darmstadt, Germany}
\author{M.~Miski-Oglu}
\affiliation{Institut f\"{u}r Kernphysik, Technische
Universit\"{a}t Darmstadt, D-64289 Darmstadt, Germany}
\author{A.~Richter}
\affiliation{Institut f\"{u}r Kernphysik, Technische
Universit\"{a}t Darmstadt, D-64289 Darmstadt, Germany}
\affiliation{ECT*, Villa Tambosi, I-38100 Villazzano (Trento),
Italy}
\author{F.~Sch\"afer}
\affiliation{Institut f\"{u}r Kernphysik, Technische
Universit\"{a}t Darmstadt, D-64289 Darmstadt, Germany}
\author{S. Tomsovic}
\affiliation{Department of Physics and Astronomy, Washington State University,
Pullman, Washington 99164-2814, USA}
\date{\today}

\begin{abstract}

Friedel oscillations of electron densities near step edges have an analog in microwave billiards.  A random plane wave model, normally only appropriate for the eigenfunctions of a purely chaotic system, can be applied and is tested for non-purely-chaotic dynamical systems with measurements on pseudo-integrable and mixed dynamics geometries.  It is found that the oscillations in the pseudo-integrable microwave cavity matches the random plane-wave modeling. Separating the chaotic from the regular states for the mixed system requires incorporating an appropriate phase space projection into the modeling in multiple ways for good agreement with experiment. 
\end{abstract}


\maketitle 
Billiards are highly suitable for understanding and discussing classically
chaotic Hamiltonian systems~\cite{Sinai70,Bunimovich74,Berry81} and their quantum counterparts~\cite{Bohigas91,Heller91}.  In recent years, they have acquired an increasing importance for their idealization of features found in systems as varied as quantum dots~\cite{Marcus92}, planetary rings~\cite{Benet00} and nuclei~\cite{Richter06}.  They also provide an ideal model for investigating residual interaction effects on many-electron ground state properties of ballistic quantum dots in the Coulomb blockade regime~\cite{Tomsovic08,Ullmo09,Kaplan08,Urbina07}.   A significant focus of quantum-classical correspondence studies has been the statistical properties of wave functions, typically for measures which are local in energy, configuration space, or both.  However, residual interaction effects motivate the investigation of measures involving both complete spatial integrations and energy summations of the squared eigenfunctions~\cite{Tomsovic08,Ullmo09}.  It was established that the dominant features are related to boundary effects, where persistent oscillations of the electronic state density are observed.  These oscillations are  known as Friedel oscillations~\cite{Friedel58} and occur regardless of the shape of the boundary.  Their investigation is currently undergoing a renaissance~\cite{Weiss07,Kovalev08,Zhang09} driven by advances in microscopy~\cite{Crommie93,Hasegawa93}.  They were first predicted in 1952~\cite{Friedel52} for the electronic density of states in metals near the Fermi level about impurity atoms, but they are clearly seen near step edges~\cite{Crommie93,Hasegawa93}, in quantum corrals~\cite{Heller94}, and in carbon nanotubes~\cite{Lee04}. 

Our focus in this paper is the measurement of Friedel oscillations in a pseudo-integrable billiard and for the chaotic states of a billiard with mixed dynamics.  They are interesting, in part, because they give us non-trivial experimental tests of random plane wave models, which normally would only be applied for the properties of purely chaotic dynamical systems (respectively, these dynamical systems lead to application of unrestricted and restricted plane wave models).  We use flat microwave cavities~\cite{Stockmannbook}, i.e.~the maximal excitation frequency is such that only one vertical mode of the electric field is excited.  The Helmholtz equation then is mathematically equivalent to the Schr\"odinger equation of the correspondingly shaped billiard.  The billiard eigenvalues are experimentally accessible as the resonance frequencies $f_\nu$ and the squared moduli of the eigenfunctions $|\psi_\nu(\vec{r})|^2$ as the electric field intensities; $\nu$ labels the resonance and $\vec{r}$ a position within the billiard.  For the resonance frequencies determination, a vector network analyzer coupled a signal into a high quality microwave billiard via one attached antenna and compared its magnitude and phase to those of a signal received at another. The electric field intensity was measured with the perturbation-body method~\cite{Sridhar92}. For this a cylindric perturber made from magnetic rubber~\cite{Bogomolny06} was placed inside the microwave billiard and moved across the entire billiard surface by means of an external guiding magnet attached to a positioning unit. According to the Maier-Slater theorem \cite{Maier52}, the perturbation body causes a shift of the resonance frequency, which is proportional to the difference of the squared electric and magnetic field strength at its location inside the cavity. We used magnetic rubber as perturber material since it does not interact with the microwave magnetic field. Thus, the intensity of the electric field strength is obtained directly from the frequency shifts.

\begin{figure}[!hbt]
\epsfig{figure=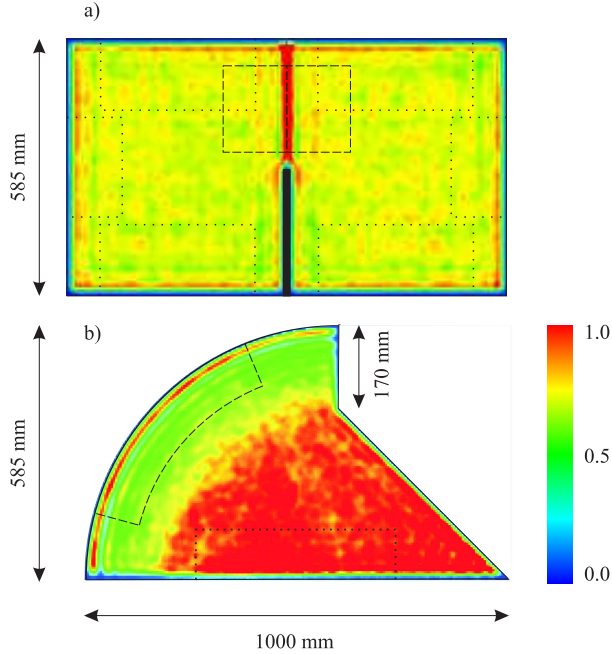,angle=0,width=8cm}
\caption{(Color online) Average intensity distribution of even eigenmodes.   In the frequency range 2--6~GHz: a) symmetric eigenmodes of the barrier billiard, and b) chaotic eigenmodes of the desymmetrized mushroom billiard.  Red color corresponds to high, blue to low intensity.  The dotted and dashed areas were used to evaluate: a) the average intensity profile at Dirichlet and Neumann boundary conditions; and b) the straight and circular boundary segments.}
\label{fig1}
\end{figure}

A first experiment was performed with the barrier billiard, i.e.~a rectangular billiard containing a barrier along the symmetry line~\cite{Hannay90}; see Fig.~\ref{fig1}a.  Its dynamics is pseudo-integrable due to trajectories hitting or missing the barrier.  The antisymmetric eigenfunctions are those of half the rectangle with Dirichlet boundary conditions.  The symmetric eigenfunctions mostly resemble chaotic wave functions in that they spread over the whole billiard surface, but roughly 20\% are superscars~\cite{Bogomolny04}; i.e.~they are localized around families of classical periodic orbits.  A total of $N=290$ symmetric intensity distributions with level numbers between $\nu=90$ and $\nu=680$ were measured for excitation frequencies above 2~GHz (see~\cite{Bogomolny06}). However, due to large noise or nearly overlapping resonances, only 239 of these could be resolved.

The second experiment measured a mixed system, a desymmetrized,
Bunimovich mushroom billiard composed of a quarter-circle joined to a
triangular stem~\cite{Bunimovich01}; see Fig.~\ref{fig1}b.  It has the interesting feature of a sharply divided phase space with one regular island and a chaotic sea.  All regular orbits reside in the quarter circle.  Their caustic radii are larger or equal to the opening width connecting the quarter-circle and stem.  Chaotic orbits are encountered throughout the billiard.  Similarly, the eigenstates may be separated into either a regular or chaotic class, with very few having an intermediate nature.  The resonance frequencies and intensity distributions of $N=239$ chaotic states were measured, again above 2~GHz, and only these were considered, thus giving a data set of size equal to that for the barrier billiard.  For a more detailed description of the experiments see~\cite{Dietz07}.
 
Both billiards have the remarkable property that their eigenfunctions can be classified and separated into two classes.  In the first class, the eigenfunctions are extremely well approximated by a semiclassical formalism based on the Einstein-Brillouin-Keller (EBK) method~\cite{EBK,Bohigas90prla,Bohigas93}.  In the second class, this is not possible.  In the barrier billiard the EBK eigenstates are those of the rectangular billiard (as mentioned above).  In the mushroom billiard, the EBK eigenstates are well approximated by those of the quarter circle billiard.  Quantum dynamical tunneling and diffraction blur the class distinction somewhat in this latter case~\cite{Dietz07} and it is interesting to know to what extent this can be ignored for the purposes of comparing to a restricted plane wave model.  For the EBK class, statistical measures involving complete spatial integrations and energy summation have been derived in~\cite{Ullmo09}.  The remaining (non-EBK) class of eigenfunctions are the symmetric ones in the barrier billiard and the chaotic ones in the mushroom billiard.  Amongst the symmetric eigenfunctions, the superscarred ones  are mainly concentrated in periodic orbit channels and their intensities are smaller, but non vanishing in the remaining part of the billiard~\cite{Aberg08}.  The chaotic eigenfunction intensity distributions are not uniform across the billiard area.  We investigate the proper application of random plane wave models for these non-EBK eigenfunctions.
   
For Dirichlet boundary conditions, the eigenstate vanishes at the edge, whereas for Neumann boundary conditions its normal derivative vanishes.  Close to the boundary, the intensity distribution is affected.  This is visible in the running total measured intensity,
\begin{equation}
\mathcal{I}(\vec r;N_2,N_1)= \sum_{\bf \nu=N_1}^{N_2} |\psi_\nu(\vec{r})|^2\, ,
\label{NE}
\end{equation}
which would be the total particle density for ultra-cold Fermi gases if $N_1=1$ and the energy associated with $N_2$ is the Fermi energy.  The summation over $\nu$ corresponds to an energy smoothing which naturally leads to a separation of this quantity into a smoothly varying secular part (associated with the Friedel oscillations) and remaining quantum fluctuations. As long as the energies associated with $N_2$ and $N_1$ span a domain at least as wide as the Thouless energy, the secular features are expected to be dominant.  Through the energy-time uncertainty principle, the Thouless energy scale corresponds to the mean time a particle takes to discover the size of the system (particle traversal time of the billiard at the Fermi velocity)~\cite{Ullmo08}.  Note that this time scale is much shorter than the Heisenberg time, which corresponds to the mean energy spacing in the quantum spectrum, and so the Thouless energy scale necessarily includes many levels.

Of special interest in~\cite{Tomsovic08,Ullmo09} was a theoretical understanding of the residual Coulomb interaction's contribution to the ground-state energy within the short-range approximation; the energy modification can be expressed in terms of single-particle eigenfunctions of closed or nearly closed quantum dots modeled as billiards.  In first-order perturbation theory, the increase of the interaction energy associated with the promotion of a particle from one orbital to another, and thus the mesoscopic fluctuations of the residual energy term and similar quantities require as an input the quantity $\mathcal{I}(\vec r;N,1)$.  Through this analogy, microwave billiards provide a means to investigate experimentally a many-body phenomenon.   Note also that the running intensity is probed directly in scanning tunneling microscopy as states below the Fermi energy contribute to the tunnel current~\cite{Fiete03}.  

Figure~\ref{fig1} shows in color scale the resulting total intensity distributions for the barrier and the mushroom billiard, respectively; each integrated $\vert\psi_\nu (\vec r)\vert^2$ was properly normalized to unity, see below.  For the barrier billiard, the total intensity of the symmetric eigenfunctions is particularly high along the symmetry line above the barrier tip.  Indeed, the eigenfunctions may be considered as the solutions of, respectively, half of the barrier billiard with Dirichlet boundary conditions along the rectangular boundary and the barrier, and Neumann boundary conditions along the opening connecting the halves. In each half, an oscillatory structure is visible, which is enhanced close to the boundary of the rectangle and the barrier. 

As noted above, only the eigenfunctions of chaotic states were taken into account in the mushroom billiard. The total intensity is considerably higher in the part which is accessible only to the chaotic orbits, namely in the stem and quarter circle part with radius equal to the opening width~\cite{Dietz07}. This can be understood in terms of the classical dynamics as its classical counterpart 
is the probability to find chaotic orbits at position $\vec r$ in the billiard. Again, close to the billiard boundary an oscillation pattern is observed. This is more clearly visible close to the circular boundary than near the straight edges.

A semiclassical expression for the total intensity's secular behavior in billiards is given by~\cite{Hoermander85}
\begin{equation}
	\mathcal{I}(\vec{r},N,1)=\frac{N_W(k_N)}{\mathcal{A}}\left(1\mp\frac{{\textrm J}_1(2k_N x)}{k_N x}\right)\, ,
	\label{friedel}
\end{equation}
where $x$ is one component of a locally defined coordinate system (see below) which measures the perpendicular distance from the boundary, $k_N$ is the wave vector modulus at frequency $f_N$, $\mathcal{A}$ is the area, and $N_W(k_N)\ \left(=\frac{\mathcal{A}k_N^2}{4\pi}\right)$ denotes the  Weyl formula leading term, i.e. the normalization is unity to leading order in $k$.  The $-$, $+$ signs refer to Dirichlet and Neumann boundary conditions, respectively, and ${\textrm J}_1(\cdot)$ is the Bessel function.  This result can be derived with a random plane-wave model normally used to simulate the eigenfunctions of a chaotic system~\cite{Berry77,Voros79b}.  Within this model the eigenfunctions near the boundary are mimicked by the superposition of a large number of plane waves~\cite{Berry02},
\begin{equation}
\psi_N(\vec r)=\frac{1}{\sqrt{N_{eff}\mathcal{A}}}\sum_{l=1}^{N_{eff}}a_l{\rm cs}(\vec k_l\hat x)\cos(\vec k_l\hat y+\varphi_l),
\label{RPWM}
\end{equation}
with random orientations of the wave vector $\vec k_l$, phases $\varphi_l$, amplitudes $a_l$, where $<a_la_{l^\prime}>=\delta_{ll^\prime}$, and fixed wave vector modulus $|\vec k_l|=k_N$. The boundary conditions are built in using the local coordinates $(\hat x,\hat y)$ along the billiard boundary, with the vectors $\hat x$ perpendicular and $\hat y$ parallel to the boundary. The function ${\rm cs}(\cdot)$ equals $\sin(\cdot)$ for Dirichlet, and $\cos(\cdot)$ for Neumann boundary conditions. In fact, here the random orientations, phases, and fixed wave vector modulus are only being used to simulate the appropriate uniform density in phase space, and not for looking at the quantum fluctuations. Squaring the
resulting wave function, integrating over the orientations, and averaging over the random phases and amplitudes yields \cite{Berry02}
\begin{equation}
\langle\vert\Psi_\nu(\vec r)\vert^2\rangle= \frac{1}{\mathcal{A}}\left[1\mp{\textrm J}_0(2k_\nu x)\right]\, , 
\label{diffriedel}
\end{equation}
where $\langle .\rangle$ denotes a local averaging over the eigenstates in a narrow frequency interval around $f_\nu$, although no finer than the Thouless energy.   The normalization is such that to leading order in $k_\nu $, its area integral equals unity~\cite{Tomsovic08,Ullmo09}.   This result is fully consistent with Eq.~(\ref{friedel}).  We stress that due to the summations, Eqs.~(\ref{friedel}), (\ref{diffriedel}) are dominated by their secular variation with frequency~\cite{Tomsovic08,Ullmo09}, thus implying that they do not depend on the nature of the classical dynamics.  This is consistent with expectations that the oscillations observed close to the boundary are due to the boundary conditions and a maximum wavelength scale defined by $k_N$, but not the billiard shape, and leads to the expectation that the random plane wave model can be applied beyond just purely chaotic dynamical situations.

\begin{figure}[!hbt]
\epsfig{figure=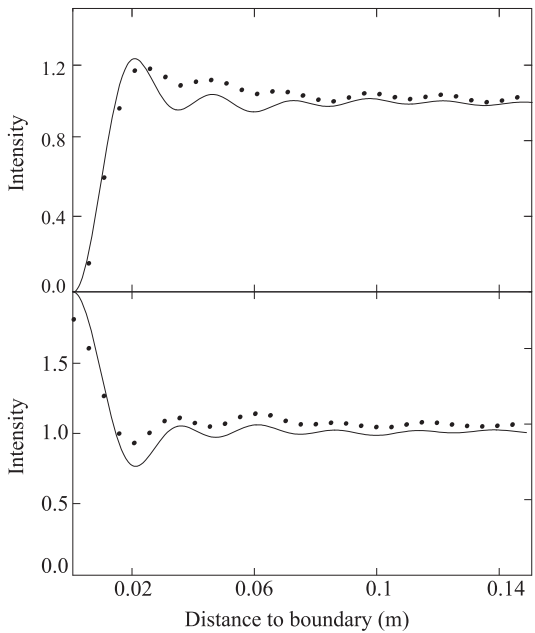,angle=0,width=7cm}
\caption{Experimental intensity profile of even eigenmodes of the barrier billiard (points) perpendicular to the Dirichlet boundary segments (upper panel) and to the Neumann boundary segment beyond the barrier (lower panel). The random plane-wave model predictions are the solid lines.}
\label{fig2}
\end{figure}

To test Eqs.~(\ref{friedel}), (\ref{diffriedel}) experimentally, the total intensity along lines of constant $x$ were determined for each frequency and summed. For the case of Dirichlet boundary conditions this was done in the barrier billiard separately for the six areas enclosed by dotted lines in Fig.~\ref{fig1}a. The upper panel of Fig.~\ref{fig2} shows the normalized average as dots. 
The solid line is obtained  by using the analytic expression for $\mathcal{I}(\vec{r},N_2,N_1)$ (see Eq. (\ref{friedel})), appropriate $(N_1,N_2)$, and dividing by $N_2-N_1+1$.  The experimental data and theory are in close agreement.  Next is the intensity distribution near the symmetry line enclosed by the dashed line in Fig.~\ref{fig1}a where the eigenfunctions obey Neumann boundary conditions.  Again the agreement is quite good, c.f.\ lower panel of Fig.~\ref{fig2}.  Note the strong enhancement of the wave function at $x=0$, which is expected.  The deviations of roughly 10~\% are attributed to the finite system size, a limited precision in the determination of the local field intensity, a few missing levels, and the number of levels taken into account.  

\begin{figure}[!hbt]
	\epsfig{figure=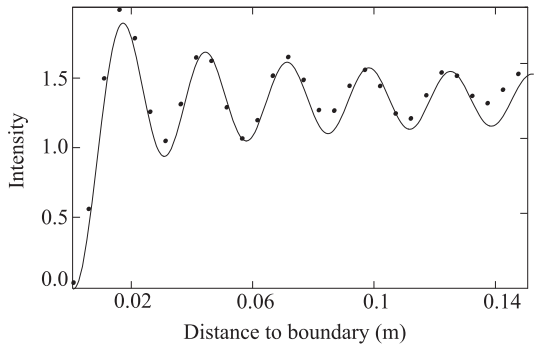,angle=0,width=7cm}
	\caption{Similar to the upper panel of Fig.~\ref{fig2} except for only 19 eigenmodes in a narrow
	frequency window 5.51--5.77~GHz.}
	\label{fig3}
\end{figure}

To test Eq.~(\ref{diffriedel}), the average intensity (six dotted areas) was measured in a range containing only 19 eigenstates.  Figure~\ref{fig3} shows the intensity profile from the interval $f \in \{5.51, 5.77\}$~GHz compared with the model prediction using the mid-point frequency.  This interval corresponds to 
roughly that deduced from the time a particle needs to travel from the billiard center to the boundary.  The oscillations are more prominent than those obtained by including a larger frequency range (c.f.\ upper panel of Fig.~\ref{fig2}) because the contributing eigenfunctions all have nearly the same wave number.  The close agreement supports the hypothesis that both $\mathcal{I}(\vec r;N_2,N_1)$ and $\langle\vert\Psi_\nu(\vec r)\vert^2\rangle$ are independent of the classical dynamics. 

The intensity distribution in the mushroom billiard was evaluated close to a straight part of the boundary (dotted line) and close to the circular boundary (dashed line).  The restriction to chaotic eigenstates injects three new facets. The least significant is that the normalization requires knowing the total number of states, 
\begin{figure}[!hbt]
        \epsfig{figure=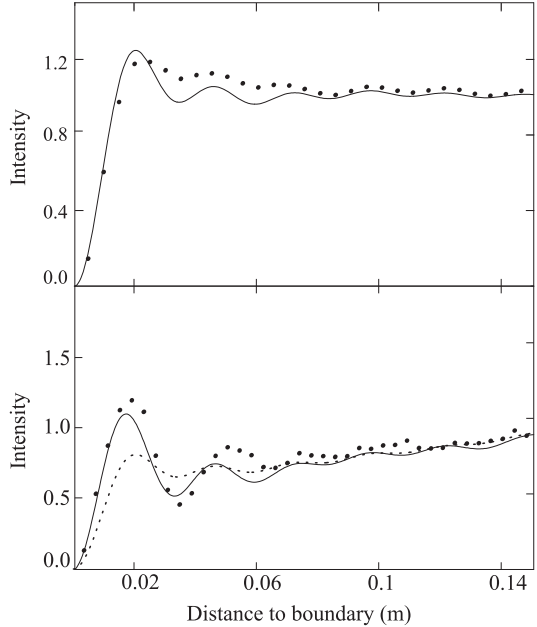,angle=0,width=7cm}
	\caption{Average intensity profile of chaotic eigenmodes (points) of the
	mushroom billiard perpendicular to its straight boundary segment
	(upper panel) and  to its circular boundary segment (lower panel) and 
        model predictions of the unrestricted and the
        restricted random plane-wave model (dashed and solid line,
        respectively).} 
        \label{fig4}
\end{figure}
not just the number of chaotic states.   In the region accessible only to chaotic orbits (see Fig.~\ref{fig1}b), the intensity is almost uniform and used to set the normalization.  The comparison of the experimental results (dots) with
the theory (solid line) obtained with Eqs.~(\ref{friedel}), (\ref{NE}) again shows good agreement (see upper panel of Fig.~\ref{fig4}). 

The second and third facets arise in the region jointly occupied by regular and chaotic eigenstates. The second one is that the average intensity decreases towards the circular boundary as $p_{\rm chaos}(x)=\frac{2}{\pi}\sin^{-1}\frac{r}{R-x}$ in the quarter ring with
inner and outer radii $r$ equal to the opening width and $R$ to the radius of the quarter circle hat, respectively \cite{Bunimovich01}. Here, $p_{\rm chaos}(x)$ is the classical probability to find a chaotic orbit within this region at a distance $x$ from the circular boundary.  This is a consequence of the phase space structure.  It was shown in~\cite{Dietz07} that the measured normalized intensity distribution Eq.~(\ref{NE}) follows this classical prediction. 

Finally, the third facet is that in the lower panel of Fig.~\ref{fig4} the measured oscillations (dots) are compared  to the analytic expression for $\mathcal{I}(\vec r;N_2,N_1)$ obtained with Eq.~(\ref{friedel}) multiplied by  $p_{\rm chaos}(x)$ shown as the dotted line.  The measured oscillations are clearly stronger.  This is due to the restriction on the orientations of waves emanating from the stem (or chaotic region).  In Eq.~(\ref{RPWM}) all orientations at the circular arc are equally probable, whereas waves from the stem are, crudely speaking, following chaotic trajectories with a maximum incidence angle with respect to the normal to the boundary of $\sin^{-1}(r/R)$.  Indeed, the classical dynamics of particles impinging the circular boundary with a higher reflection angle is regular, as they never enter the stem.  This is accounted for by restricting in Eq.~(\ref{RPWM}) the random orientations of the wave vector at the boundary to the angle interval $\left[-\sin^{-1}(r/R),\sin^{-1}(r/R)\right]$. Restricted plane wave models have been applied before in~\cite{Baecker02} to describe distributions in billiard systems of mixed dynamics.  Analytic expressions for the restricted and projected $\mathcal{I}(\vec r;N_2,N_1)$  and $\langle\vert\Psi_\nu(\vec r)\vert^2\rangle$ are cumbersome and here calculated numerically.  The result is the solid line in the lower panel of Fig.~\ref{fig4}. It provides a much better description of the oscillations than the unrestricted version.  The remaining deviations above $x=0.08$ are attributed to dynamic tunneling across the quarter circular border defining the smallest possible caustic of the regular orbits. It is argued in~\cite{Dietz07} that this phenomenon causes a distortion of the wave functions along the border and quantified these deviations. Thus the classification of modes as chaotic and regular ones is correct only asymptotically in the semiclassical limit even in mushroom billiards with the clearly separated phase space. As a consequence some of the considered chaotic modes contain a regular admixture and this causes the observed deviation.

In summary, we detected Friedel oscillations of the total intensity near the boundary of a barrier billiard and a desymmetrized mushroom billiard. We showed that the oscillations can be understood theoretically both for Dirichlet and Neumann boundary conditions and that the features of the oscillations do not depend on the system dynamics.  However, when a restricted set of modes is considered, system specific properties have to be incorporated which restrict the wave orientations.  Interestingly, an enhanced intensity is found, if not all angles of incidence to the boundary are allowed.  We propose that this enhancement can be regarded as a dynamical localization effect in mesoscopic systems from the perspective that part of the supposedly available 
momentum space is not being accessed.  This work shows that although microwave billiards allow only for the measurement of single particle wave functions, the measured data can also be used to reconstruct properties of many-body systems.

This work was supported by the DFG within SFB~634.  F.~S.\ is grateful for financial support from Deutsche Telekom Foundation.  S.T.~gratefully acknowledges support from  the U.S.~NSF Grant  PHY-0855337, and the Max-Planck-Institut f\"ur Physik komplexer Systeme.

\end{document}